\documentstyle[prl,aps,multicol]{revtex}
\input epsfig.sty
\newcommand{\be}{\begin{equation}}
\newcommand{\la}{\label}
\newcommand{\ee}{\end{equation}}
\newcommand{\bea}{\begin{eqnarray}}
\newcommand{\eea}{\end{eqnarray}}
\newcommand{\rp}{\dot{R}}
\begin{document}

\title{A dynamical chiral bag model}
\author{K.~Colanero and M.-C.~Chu}
\address{Department of Physics, The Chinese University of Hong Kong, 
Shatin, N.T., Hong Kong.}
\maketitle
\begin{abstract}
We study a dynamical chiral bag model, in which massless fermions are
confined within an impenetrable but movable bag coupled to meson fields.
The self-consistent motion of the bag is obtained by solving the 
equations of motion exactly assuming spherical symmetry.
When the bag interacts with an external meson wave  we find 
three different kinds of resonances: {\it fermionic},
{\it geometric}, and $\sigma$-resonances.
We discuss the phenomenological implications of our results.
\end{abstract}
\pacs{PACS number(s): 12.39.Ba, 12.39.Fe, 14.20.-c}  

\begin{multicols}{2}

\narrowtext

\section{Introduction}
The MIT Bag model \cite{MIT} and its chirally invariant versions, 
such as the Chiral Bag model \cite{chodos,chbag} and the Cloudy Bag 
model \cite{cloudy}, continue to be useful
tools in the study of the physics of the nucleon and other baryons. 
They have also been used extensively in the discussion of various 
phenomena ranging from strange stars \cite{sstar} to ultra-relativistic 
heavy-ion collisions \cite{rhic}, even though these often involve 
situations of high density/temperature where the applicability of the 
models is doubtful. 

In most of the bag model studies so far,  because of its simplicity,
a static spherical bag is assumed.  The few notable exceptions,
which allowed for the possibility of
a dynamical bag boundary, focused mainly on reproducing the correct 
phenomenological parity order of the low-lying states of the nucleon,  
although several approximations and modifications to the theory had to 
be employed.  For example, Rebbi and DeGrand \cite{rebbi}
studied a bosonic bag and quantized the full system with perturbation 
theory in the limit of small spherical oscillations. The authors in 
Ref.~\cite{kuti,brown,meissner}
considered a fermionic bag with a surface tension, as well as the volume 
energy, and quantized only 
the motion of the bag boundary in the adiabatic approximation. Nogami and 
Tomio \cite{nogami}
also quantized the motion of the boundary, but used the adiabatic 
approximation only for the mesons. Although these works gave a reasonable 
ordering of the low-lying states of the nucleon, the more fundamental
question of whether it is consistent and feasible to use a dynamical bag
to model hadrons was not addressed.  That is the motivation of 
the present work. 

In a previous paper \cite{colchu3} 
we proved that the original MIT bag model with massless quarks 
admits only one classical solution other than 
the static one, namely, a bag constantly expanding at the speed of light.
We thus concluded that an additional field, such as the mesons in the
Chiral Bag model, is needed to have a consistent and nontrivial dynamical
bag model of hadrons. In this paper we implement a method that allows us 
to find the classical solutions of a spherically 
symmetric chiral bag for any motion of the bag radius. 
In particular we look for the self-consistent solution of the full
theory without any approximation, in which the 
motion of the bag surface is determined by the conservation of the total 
energy.  We can thus study the full nonlinear features of the model.

This paper is organized as follows. We first show the
method we use to solve the problem. We then discuss the resonances
found with a driven bag motion. In the third section the problem of the 
self-consistent surface motion is addressed, and we discuss the results
obtained with different incoming meson waves. We finally summarize our
results and discuss their phenomenological implications.
The appendix provides more details about the method of solution.

\section{Method of solution}

The Lagrangian of the system we study is \cite{chodos}:

\bea
{\cal L} = {1 \over 2} \left\{\hspace{-0.2 cm} \begin{array}{l}
\\
\\
\end{array}
\left[i\left(\bar\psi \gamma^\mu 
\partial_\mu \psi - 
(\partial_\mu \bar\psi) \gamma^\mu \psi\right)- B\right] \theta_V (x) - \right. \nonumber \\
\left. {1\over f_\pi} \bar\psi (\sigma +i \vec{\tau} \cdot \vec{\pi} \gamma_5)
\psi \Delta_s +\partial_\mu \sigma \partial^\mu \sigma +
\partial_\mu \vec{\pi} \cdot \partial^\mu \vec{\pi}\right\} \ \ ,
\la{lagct}
\eea
where $\theta_V (x)$ is $1$ inside the bag and $0$ outside and 

\be
{\partial \theta_V \over \partial x^\mu} = n_\mu \Delta_s \; ,
\la{dthetadmu}
\ee
$\Delta_s$ being the surface delta-function.
From it we derive the following Euler-Lagrange equations of motion:

\be
\gamma^\mu \partial_\mu \psi = 0 \hspace{1 cm} \mbox{inside the bag,}
\la{eqmn1}
\ee

\be
i \gamma^\mu n_\mu \psi = {1\over f_\pi} (\sigma + i \vec{\tau} \cdot 
\vec{\pi} 
\gamma_5 ) \psi \hspace{1 cm} \mbox{on the bag surface,}
\la{eqmn2}
\ee

\be
\partial_\mu \partial^\mu \sigma=-{1\over 2 f_\pi} \bar\psi \psi \Delta_s 
\: ,
\la{eqmn3}
\ee

\be
\partial_\mu \partial^\mu \vec{\pi}= -{1\over 2 f_\pi} i \bar\psi \gamma_5
\vec{\tau} \psi \Delta_s \: .
\la{eqmn4}
\ee
It is possible to look for spherically symmetric solutions of the above 
equations \cite{chodos} by writing:
\be
\psi = \left ( \begin{array}{c}
	g(t,r) \\ -i \vec{\sigma} \cdot \hat{r} f(t,r)
	\end{array} \right ) v \: ,
\la{dirspin}
\ee
\be
\sigma=\sigma(t,r) \: ,
\la{sigma}
\ee
\be
\vec{\pi}=\pi(t,r) \hat{r} \: ,
\la{pi}
\ee
where $v$ includes the spin and isospin parts and can be written as
\be
v={1\over 2}\left(|\uparrow , d> - |\downarrow , u>\right) \; .
\la{v}
\ee
The arrows indicate the spins while $u$ and $d$ the up and down flavors of 
the quarks, and $v$ satisfies 
\be
(\vec{\sigma} + \vec{\tau}) v = 0 \; .
\la{hedgecond}
\ee
In Eqs.~\ref{dirspin}~and~\ref{hedgecond}, $\vec{\sigma}$ are the 
three Pauli matrices and should not be confused with the field 
$\sigma(t,r)$.
Eq.~\ref{hedgecond} ensures that the RHS of Eqs.~\ref{eqmn2}~and~\ref{eqmn4} 
are spherically symmetric and causes $\vec{\pi}$ to be radially 
directed, 
as in Eq.~\ref{pi}. This kind of solutions is hence called hedgehog 
solutions \cite{chodos,bhaduri}.

For a static bag an analytic solution is known \cite{chodos,bhaduri}, which 
represents a stationary 
fermion field coupled at the surface of the bag to time-independent 
$\sigma$ and $\pi$-fields. Our goal is to find the hedgehog solution 
for any spherically symmetric motion of the bag's surface. 

Substituting Eq.~\ref{dirspin} for $\psi$ in Eq.~\ref{eqmn1} we obtain

\be
i{\partial f\over \partial t} = {\partial g \over \partial r} \ \ ,
\la{eqmn1rd}
\ee
\be
-i{\partial g\over \partial t} = {\partial f\over \partial r}+{2\over 
r} f \: .
\la{eqmn2rd}
\ee
It is not difficult to verify \cite{colchu3} that the general solution of 
Eqs.~\ref{eqmn1rd},~\ref{eqmn2rd} has the form:

\be
g(t,r) = {1 \over r}\left[Q^\prime (t-r) - Q^\prime (t+r) \right] \ \ , 
\la{g(t,r)}
\ee
\bea
f(t,r) = {i\over r}\left\{\hspace{-0.2 cm} \begin{array}{l}
\\
\\
\end{array}
Q^\prime (t-r) + Q^\prime (t+r) \right. + \nonumber \\
\left. {1\over r} 
\left[Q(t-r)-Q(t+r)\right]\right\} \: ,
\la{f(t,r)}
\eea
where $Q(z)$ is an arbitrary function.
In spherical coordinates Eqs.~\ref{eqmn3}~and~\ref{eqmn4} become

\be
{\partial^2 \sigma \over \partial t^2} - {\partial^2 \sigma \over 
\partial r^2} -
{2\over r} {\partial \sigma \over \partial r} = -{1\over 2f_\pi} 
\left[ g^* g - f^* f \right] \delta(R-r) \: ,
\la{eqmn3rd}
\ee

\be
{\partial^2 \pi \over \partial t^2} - {\partial^2 \pi \over \partial r^2} 
-
{2\over r} {\partial \pi \over \partial r} + {2\over r^2} \pi
= {1\over 2f_\pi} \left[g^* f+gf^*\right] \delta(R-r) \: .
\la{eqmn4rd}
\ee
For $r \ne R$, we notice that $\sigma(t,r)$ obeys the equation 
of a free $s$-wave while
$\pi(t,r)$, being the radial part of the vector $\vec{\pi}$, satisfies 
the equation of a free $p$-wave.
Hence we can look for a solution of the form

\be
\sigma(t,r)= \sigma_{\rm in}(t,r) \theta(R-r) + \sigma_{\rm out}(t,r) 
\left[1-\theta(R-r)\right] \: ,
\la{sigmainout}
\ee
\be
\pi(t,r)= \pi_{\rm in}(t,r) \theta(R-r) + \pi_{\rm out}(t,r) \left[1-
\theta(R-r)\right] \: , 
\la{piinout}
\ee
where the fields inside and outside of the bag can be written accordingly 
as

\be
\sigma_{\rm in}(t,r)={1 \over r}\left[\Sigma_{\rm in}(t-r) -
\Sigma_{\rm in}(t+r) \right] + 
\sigma_{\rm 0,in}(r) \: ,
\la{sigmain}
\ee
\be
\sigma_{\rm out}(t,r)={1 \over r}\left[\Sigma_{\rm out-}(t-r) - 
\Sigma_{\rm out+}(t+r) \right] + 
\sigma_{\rm 0, out}(r) \: ,
\la{sigmaout}
\ee
\bea
\pi_{\rm in}(t,r)={1 \over r}\left\{\hspace{-0.2 cm} \begin{array}{l}
\\
\\
\end{array}
\Pi^\prime_{\rm in}(t-r)+
\Pi^\prime_{\rm in}(t+r)\right. + \nonumber \\
\left. {1\over r}\left[\Pi_{\rm in}(t-r)-\Pi_{\rm 
in}(t+r)
\right] \right\} + \pi_{\rm 0, in}(r) \: ,
\la{piin}
\eea
\bea
\pi_{\rm out}(t,r)={1 \over r}\left\{\hspace{-0.2 cm} \begin{array}{l}
\\
\\
\end{array}
\Pi^\prime_{\rm out-}(t-r)+
\Pi^\prime_{\rm out+}(t+r)\right. + \nonumber \\
\left. {1\over r}\left[\Pi_{\rm out-}(t-r)-\Pi_{\rm out+}(t+r)\right] \right\} + 
\pi_{\rm 0, out}(r) \: ,
\la{piout}
\eea
where $\Sigma_{\rm out+}$, $\Sigma_{\rm out-}$, $\Sigma_{\rm in}$,
$\Pi_{\rm in}$, $\Pi_{\rm out+}$ and $\Pi_{\rm out-}$ are arbitrary 
functions. Notice that $\Sigma_{\rm out+}$ and $\Sigma_{\rm out-}$ are 
in general different functions as are also 
$\Pi_{\rm out+}$ and $\Pi_{\rm out-}$.
Here, the time-independent terms, $\sigma_{\rm 0,in}(r)$, 
$\sigma_{\rm 0, out}(r)$, $\pi_{\rm 0, in}(r)$ and $\pi_{\rm 0, out}(r)$ 
are 
the static-bag solutions given by \cite{chodos,bhaduri}

\begin{eqnarray*}
\sigma_{\rm 0,in}(r) & = & g_0 \ \ , \\
\sigma_{\rm 0, out}(r) & = & g_0 + \alpha R_0^2 \left({1\over R_0} - 
{1\over r}\right) \ \ , \\
\pi_{\rm 0, in}(r) & = & -{\beta\over 3} r \ \ , \\
\pi_{\rm 0, out}(r) & = & -{\beta\over 3} {R_0^3\over r^2} \: .
\end{eqnarray*}

Substituting Eqs.~\ref{sigmainout}~and~\ref{piinout} in 
Eqs.~\ref{eqmn3rd}~and~\ref{eqmn4rd} and requiring the
%
%
continuity of $\sigma(t,r)$ and $\pi(t,r)$ at $r=R$, 
we finally obtain two relations that can be viewed as 
boundary conditions for the fields $g$, $f$, $\sigma_{\rm in}$, 
$\sigma_{\rm out}$, $\pi_{\rm in}$ and $\pi_{\rm out}$:

\bea
\dot{R} \left({\partial \sigma_{\rm in}\over \partial t}-{\partial 
\sigma_{\rm out}\over \partial t}\right) +
\left({\partial \sigma_{\rm in}\over \partial r}-{\partial 
\sigma_{\rm out}
\over  \partial r}\right)= \nonumber \\
-{1\over 2f_\pi} \left(g^* g - f^* f\right) 
\hspace{1 cm} \mbox{at} \hspace{0.3 cm} r=R \ ,
\la{bndsigma}
\eea

\bea
\dot{R} \left({\partial \pi_{\rm in}\over \partial t}-
{\partial \pi_{\rm out} \over \partial t}\right) +
\left({\partial \pi_{\rm in}\over \partial r}-
{\partial \pi_{\rm out}\over \partial r}\right) = \nonumber \\
{1\over 2f_\pi} \left(g^* f+gf^*\right) \hspace{1 cm} \mbox{at} 
\hspace{0.3 cm} r=R \ .
\la{bndpi}
\eea

From Eq.~\ref{eqmn2} we can express $\sigma(t,R)$ and $\pi(t,R)$ in terms
of $g(t,R)$ and $f(t,R)$ (see Appendix) and use  
Eqs.~\ref{bndsigma}~and~\ref{bndpi} 
to find $g(t,r)$, $f(t,r)$, {\it i.e.}~$Q_{\rm re}(z)$, and $Q_{\rm im}(z)$ 
(see Eqs.~\ref{g(t,r)},~\ref{f(t,r)})
with the null-lines method \cite{cole,colchu2}.

From the point of view of the null-lines method the unknowns in 
Eqs.~\ref{bndsigma}~and~\ref{bndpi},
as long as $|\rp| \leq 1$ \cite{colchu2}, are $Q_{\rm re}(t+R)$, 
$Q_{\rm im}(t+R)$, $\Sigma_{\rm in}(t+R)$, $\Pi_{\rm in}(t+R)$, 
$\Sigma_{\rm out-}(t-R)$, and $\Pi_{\rm out-}(t-R)$. 
Furthermore, the latter four are fixed once $Q_{\rm re}(t+R)$ and
$Q_{\rm im}(t+R)$ are known, using

\be
\Sigma_{\rm in}(t+R) = \Sigma_{\rm in}(t-R) + R g_0 - R \sigma (t,R) \ ,
\la{sin+}
\ee
\bea
\Sigma_{\rm out-}(t-R) =  \Sigma_{\rm out+}(t+R) + \alpha R_0^2 - \nonumber \\
R \left( g_0+\alpha R_0 \right) + R \sigma (t,R) \: ,
\la{sout-}
\eea

\bea
\Pi^\prime_{\rm in}(t+R) = {1\over R} \left[\Pi_{\rm in}(t+R) - 
\Pi_{\rm in}(t-R)\right] + \nonumber \\
\Pi^\prime_{\rm in}(t-R) +
{\beta\over 3} R^2 + R \pi (t,R) \ ,
\la{pin+}
\eea
\bea
\Pi^\prime_{\rm out-}(t-R) = {1\over R}\left[\Pi_{\rm out+}(t+R)- 
\Pi_{\rm out-}(t-R) \right] - \nonumber \\
 \Pi^\prime_{\rm out+}(t+R)  +
{\beta\over 3}{R^3_0\over R} + R \pi (t,R) \: .
\la{pout-}
\eea
Here for convenience we have not written explicitly the dependence on 
$Q_{\rm re}(t+R)$ and $Q_{\rm im}(t+R)$ which are hidden in
$\sigma (t,R)$ and $\pi (t,R)$.

We still need to solve Eqs.~\ref{pin+}~and~\ref{pout-}.
This can be done numerically either by simply replacing 
$\Pi^\prime_{\rm in}(t+R)$ and $\Pi^\prime_{\rm out-}(t-R)$ with their 
finite-difference counterparts or by integrating between $z-dz$ and $z$, 
where $z=t+R$ and $z=t-R$ respectively for the
first and second equations, and  by approximating all quantities other 
than $\Pi_{\rm in}(t+R)$ and $\Pi_{\rm out-}(t-R)$ 
as being constant in this infinitesimal interval. The numerical results 
turn out to be slightly more accurate with the second method. 
To solve Eqs.~\ref{bndsigma}~and~\ref{bndpi} we used a fourth-order 
Runge-Kutta algorithm.

\section{Resonances with a driven bag motion}

\begin{figure}
\psfig{file=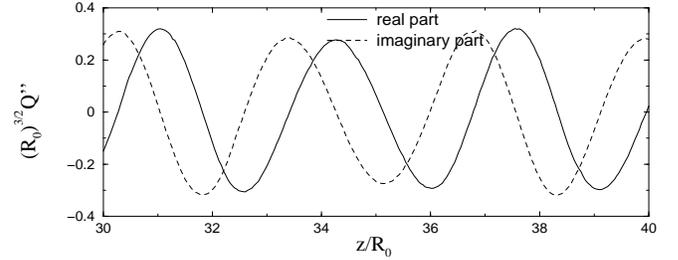,angle=-90,width=8.5 cm}

\caption{Time evolution of the real and imaginary 
parts of  $Q^{\prime\prime}(z)$ for driven 
surface oscillations, with $\nu=1/R_0$ and $\epsilon=0.05R_0$. }
\label{fig2}
\end{figure}

With the method discussed in the previous section we first computed the 
solution for a static bag and then for a slowly moving one. We 
verified that the norm
of the fermion field is conserved and that our numerical method is
accurate up to the second derivative of $Q(z)$
for a bag of initial radius $R_0=1$ fm and $f_\pi =1$ fm$^{-1}$. With these 
parameters the static chiral bag is similar
to the MIT bag, with an almost constant $\sigma(r)$ and a very 
small $\pi(r)$. In Fig.~\ref{fig2} we plotted 
the second derivative of $Q(z)$ in order to show the quality of the 
numerical solution. 
All the results presented below are obtained with 
$R_0=1$ fm and $f_\pi =1$ fm$^{-1}$, which is a representative
set of parameters for showing the qualitative features of a
dynamical chiral bag model. 

Since we are particularly interested in the behaviour of the fields under 
the effect of the motion of the boundary, 
we first study the chiral bag with an imposed surface motion. 
Subjecting the bag boundary to a sinusoidal motion,
$R(t) = R_0+\epsilon[\cos (\nu t) -1]$,
we found three different kinds of resonances: i) the {\it fermionic} 
resonances, which are excited when the oscillation 
frequencies are close to the difference between two static-bag eigen-energies, 
$\nu \simeq E_n-E_k$;  ii) the {\it geometric} $\sigma$ resonances, 
for $\nu \simeq n \pi /R_0$; iii) the {\it parametric} $\sigma$ resonances,
for $\nu \simeq (2n+1)\pi/(2R_0)$, where $n$ is an integer.

\begin{figure}
\psfig{file=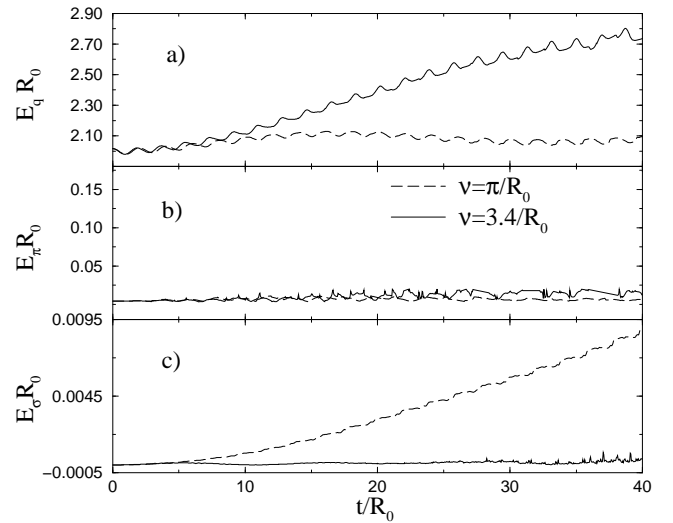,angle=0,width=8.5 cm}

\caption{Time evolution of the energy for driven surface oscillations 
at the $\nu=\pi/R_0$ (dashed lines) and 
$\nu=E_2-E_1$ (solid lines) resonances with $\epsilon=0.01R_0$.
a) Energy of the fermion field. b) Energy of the $\pi$-field. 
c) Energy of the $\sigma$-field.}
\label{fig3}
\end{figure}

The origin of the fermionic resonances at $\nu \simeq E_n - E_k$ is similar to 
those found for a Schr\"{o}dinger particle in an oscillating cavity 
\cite{colchu1}. The difference here is that the fermions 
cannot really be excited to the upper static-cavity level because the 
upper level is associated with different static 
pion fields which cannot be produced by the boundary motion. 
However, since with our choice of parameters 
$\sigma(r)$ and $\pi(r)$ change little for different static solutions, 
the system still gets excited for
oscillation frequencies close to the static energy gaps. The  
smaller $f_\pi$ is, these resonance frequencies deviate more from $E_n - E_k$.
As an example, we show for $\nu=3.4/R_0 \simeq E_2-E_1$
the time-dependence of the 
energies of the fermion and the meson fields 
in Fig.~\ref{fig3}a, b and c respectively.
It is interesting to note that neither the $\sigma$ nor the 
$\pi$-fields gain considerable energy.

For $\nu = n \pi/R_0$ we found resonances involving the $\sigma$-field. 
As can be seen in Fig.~\ref{fig3} the energy 
of the $\sigma$-field increases 
remarkably, while the energies of the fermion and the $\pi$-field change 
little. These resonances may be
considered {\it geometric}, since the resonance frequencies are related 
to the time it takes for the wave components of $\sigma(t,r)$, 
{\it i.e.}, $\Sigma_{\rm in} (t-r)$
and $\Sigma_{\rm in} (t+r)$, to travel from the boundary of the bag 
to its center and back again. 
It has been shown \cite{wai} that {\it p-waves} in an oscillating 
spherical cavity also manifest resonances at
$\nu = n \pi/R_0$, and so it is somewhat surprising that here the 
energies of the fermion and $\pi$-fields are 
little affected.  The strongly nonlinear 
interaction at the bag boundary seems to damp out the
resonant evolution. The energy of the fermions actually shows some
resonant behaviour, but this is probably mainly 
due to the fact that the driving frequency is close to $E_2 -E_1$.

\begin{figure}
\psfig{file=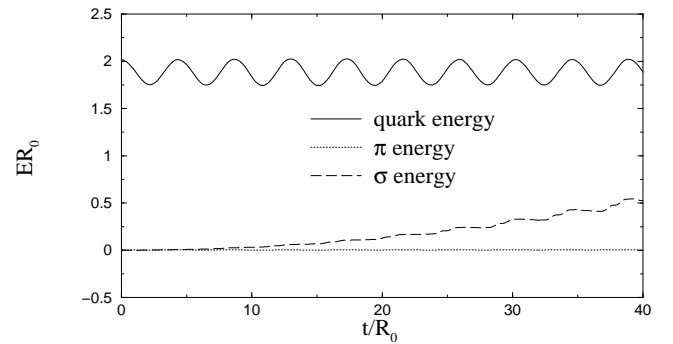,angle=-90,width=8.5 cm}

\caption{Time evolution of the energy for driven surface oscillations 
at the $\nu=\pi/(2R_{\rm av})$ resonance with $\epsilon=0.08R_0$. 
($R_{\rm av}=R_0+\epsilon$)}
\label{fig4}
\end{figure}

\begin{figure}
\psfig{file=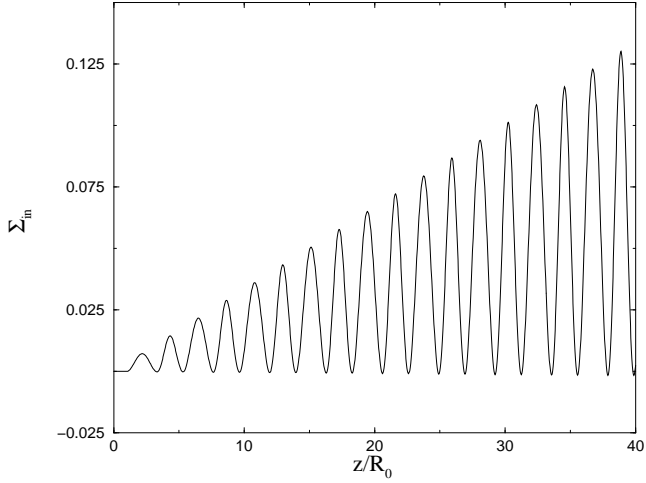,angle=-90,width=8.5 cm}

\caption{The function $\Sigma_{\rm in}(z)$ for driven surface oscillations 
at the $\nu=\pi/(2R_{\rm av})$ resonance with 
$\epsilon=0.08R_0$. The function clearly contains a periodic contribution with
$T \simeq 2/R_0$.}
\label{fig12}
\end{figure}

The third kind of resonances we found is a peculiar feature of the system 
under analysis. As we can see from 
Fig.~\ref{fig4}, it involves mainly the $\sigma$-field.
Note that $\Sigma_{\rm in} (z)$ (Fig.~\ref{fig12}) has an almost 
periodic dependence and the period is about
half that of the oscillating bag. In other words the 
bag surface, oscillating at frequencies $\nu = (2n+1)\pi/(2R_0)$, 
acts as a source for a $\sigma$-field with frequencies $(2n+1) 
\pi/ R_0$. Such frequencies are obviously 
resonant with the cavity so that the $\sigma$ field is resonantly-enhanced.
The occurence of this kind of resonances can be understood 
in the following way. 
Since the fermion field is not excited we can approximate its value 
at the bag boundary with its static-bag expression
\be
g(t,R(t)) = N \exp{(-iEt)} j_0(ER(t)) \; ,
\la{gst}
\ee
\be
f(t,R(t)) = -N \exp{(-iEt)} j_1(ER(t)) \; .
\la{fst}
\ee
From 
Eqs.~\ref{bnddirspinreim1},~\ref{bnddirspinreim2},~\ref{bnddirspinreim3},
\ref{bnddirspinreim4}, we obtain after some manipulations 
\be
\pi(t,R)=f_{\pi}{f_{\rm re}^2(t,R)+f_{\rm im}^2(t,R)-g_{\rm re}^2(t,R)-
g_{\rm im}^2(t,R)\over 
f_{\rm re}^2(t,R)+f_{\rm im}^2(t,R)+g_{\rm re}^2(t,R)+g_{\rm im}^2(t,R)} ,
\la{pioq}
\ee
\be
\sigma(t,R)=-2f_{\pi}{f_{\rm re}(t,R) g_{\rm re}(t,R)+f_{\rm im}(t,R) 
g_{\rm im}(t,R)\over 
f_{\rm re}^2(t,R)+f_{\rm im}^2(t,R)+g_{\rm re}^2(t,R)+g_{\rm im}^2(t,R)} .
\la{sigmaoq}
\ee
One can see that the dependence on $\exp{(-iEt)}$ cancels out in our
approximation and the whole expressions become
periodic functions with period $T=2\pi/\nu$.
The Fourier expansion of such functions involves all multiple 
frequencies of $\nu$ and, in the case of 
$\nu=(2n+1)\pi/(2R_0)$, its even multiples are also integral multiples of 
$\pi/R_0$. It is then evident that 
the expressions for $\sigma(R)$ and $\pi(R)$ contain terms in resonance 
with the cavity. However, it is surprising 
that for oscillation amplitudes $\epsilon > 0.005R_0$ the frequency 
$2 \nu$ for $\sigma(R)$ becomes the dominating one
even before the first bag oscillation is completed. In 
Fig.~\ref{fig12} we can see clearly how the amplitude
of $\Sigma_{\rm in}(z)$ increases with each bag oscillation.
Although the expression for $\pi(R)$ also contains terms with frequency 
being integral multiples of $\pi/R_0$, we observe no resonant 
behaviour for the $\pi$-field, which is consistent with the previous 
observation that this field is not excited for $\nu = n \pi/R_0$.

\section{Self-consistent surface motion}

\subsection{Equation for the radius and non-resonant interaction}

Our main interest in this work is to study the behaviour of the fields 
and the bag's surface when perturbed from their 
static-bag states by, for example, an incoming pion wave. To this end we 
need to find the self-consistent dynamics of the bag
surface and fields. 

We notice that the velocity of points on the bag surface does not
appear in the Lagrangian, Eq.~\ref{lagct}, and hence we cannot derive 
from it an equation of motion for
the bag radius \cite{rebbi,kuti}. The motion of the bag however is 
constrained by the conservation of the total energy.
Let us consider the energy-momentum tensor
\bea
T^{\mu\nu}=-g^{\mu\nu}{\cal L}+{i\over 
2}\left[\bar\psi\gamma^\mu\partial^\nu\psi-\partial^\nu\bar\psi\gamma^\mu
\psi\right]\theta_V + \nonumber \\
\partial^\mu\sigma\partial^\nu\sigma+\partial^\mu\vec{\pi}\cdot
\partial^\nu\vec{\pi} \; .
\la{tmunu}
\eea
The conservation of energy and momentum requires $\partial_\mu 
T^{\mu\nu}=0$, from which, by using
the equations of motion 
Eqs.~\ref{eqmn1},~\ref{eqmn2},~\ref{eqmn3},~\ref{eqmn4} and after some 
algebra, we derive
\be
B n^\nu={1\over 
2f_\pi}\partial^\nu[\bar\psi(\sigma+i\vec{\tau}\cdot\vec{\pi}\gamma_5)\psi] 
\hspace{1 cm} 
\mbox{on the surface}\; .
\la{bnnu1}
\ee
For spherically symmetric solutions $n^\nu \equiv (\dot{R} , \hat{r})$, 
and the above equation can be written as
\be
B \dot{R}={1\over 2f_\pi} \left\{ {\partial\over \partial 
t}\left[\bar\psi(\sigma+i\vec{\tau}\cdot\hat{r}\pi\gamma_5)\psi\right]
\right\}_{r=R} \; ,
\la{brdot1}
\ee
\be
B ={1\over 2f_\pi} \left\{ {\partial\over \partial r}
\left[\bar\psi(\sigma+i\vec{\tau}\cdot\hat{r}\pi\gamma_5)\psi\right]
\right\}_{r=R} \; .
\la{bhatr1}
\ee
In the static case Eq.~\ref{brdot1} is identically satisfied because 
$\dot{R}=0$ and 
$\bar\psi(\sigma+i\vec{\tau}\cdot\vec{\pi}\gamma_5)\psi$ is time 
independent, and we could use Eq.~\ref{bhatr1}
to derive $B$. However, the RHS of Eq.~\ref{bhatr1} is an ambiguous 
expression, because it involves the derivatives of
$\sigma$ and $\pi$ at the boundary which are discontinuous. To overcome 
this difficulty we use the fact 
that $T^{\mu\nu}$ can also be written as \cite{chodos}
\be
T^{\mu\nu}=T^{\mu\nu}_{\rm in} \theta_V + T^{\mu\nu}_{\rm out} 
(1-\theta_V) \ ,
\la{tmunu2}
\ee
because the surface term is zero along the trajectories of motion. Since 
$\partial_\mu T^{\mu\nu}_{\rm in}=0$ and 
$\partial_\mu T^{\mu\nu}_{\rm out}=0$, the conservation condition for 
energy and momentum becomes
\be
n_\mu T^{\mu\nu}_{\rm in} = n_\mu T^{\mu\nu}_{\rm out} \hspace{1 cm} 
\mbox{on the surface} \; .
\la{tmunuintmunuout}
\ee
Again using the equations of motion we obtain
\be
\begin{array}{l}
n_\mu T^{\mu\nu}_{\rm in}-n_\mu T^{\mu\nu}_{\rm out} = n^\nu (B-D(t))+ \\
\\
{1\over 2f_\pi}\partial^\nu\left[\bar\psi\left(\sigma_{\rm av}+i\vec{\tau}
\cdot\vec{\pi}_{\rm av}\gamma_5\right)\psi\right] \\
\\
-n_\mu \partial^\mu\sigma_{\rm in} \partial^\nu\sigma_{\rm out} + n_\mu 
\partial^\mu\sigma_{\rm out} \partial^\nu\sigma_{\rm in} \nonumber \\
\\
-n_\mu \partial^\mu\pi_{\rm in} \partial^\nu\pi_{\rm out} + n_\mu 
\partial^\mu\pi_{\rm out} \partial^\nu\pi_{\rm in} = 0 \ \ ,
\end{array}
\la{tmunuintmunuout1}
\ee
with
\be
\sigma_{\rm av}\equiv{1\over 2}\left(\sigma_{\rm in}+\sigma_{\rm out}\right) 
\; ,
\la{sigmav}
\ee
\be
\pi_{\rm av}\equiv{1\over 2}\left(\pi_{\rm in}+\pi_{\rm out}\right) \ \ ,
\la{piav}
\ee
\be
D(t) \equiv {1 \over 2}\left[\left(\partial_\rho \sigma_{\rm in}\right)^2+
\left(\partial_\rho \vec{\pi}_{\rm in}\right)^2-
\left(\partial_\rho \sigma_{\rm out}\right)^2-\left(\partial_\rho 
\vec{\pi}_{\rm out}\right)^2\right]  ,
\ee
and all the functions in the above expressions are evaluated at the 
surface of the bag.

Eq.~\ref{tmunuintmunuout1} is well defined, and since the spatial 
part of $n^\nu$ in the case of spherical symmetry
is $\hat{r}$, it can be used to calculate the bag constant $B$ as
\bea
B=D(t)+{1\over 2f_\pi}{\partial \over \partial 
r}\left[\bar\psi\left(\sigma_{\rm av}+i\vec{\tau}\cdot\vec{\pi}_{\rm av}
\gamma_5\right)\psi\right] \nonumber \\
-n_\mu \partial^\mu\sigma_{\rm in} {\partial\sigma_{\rm out} \over 
\partial r} + n_\mu \partial^\mu\sigma_{\rm out} 
{\partial\sigma_{\rm in} \over \partial r} - \nonumber \\
n_\mu \partial^\mu\pi_{\rm in}
{\partial\pi_{\rm out} \over \partial r} + 
n_\mu \partial^\mu\pi_{\rm out} {\partial\pi_{\rm in} \over \partial r} \ ,
\la{bagconstant}
\eea
where the static-bag solution has to be used. With the static hedgehog 
solution the terms involving products of
fields inside and outside the bag actually cancel out each other.
Such a value of $B$ ensures that the continuity equation for the linear 
momentum is satisfied. 
Choosing $\nu=0$ from Eq.~\ref{tmunuintmunuout1} we can derive an 
equation for $\dot{R}$:
\bea
\dot{R}={-1\over B-D(t)}\left\{ {1\over 2f_\pi}{\partial \over \partial 
t}\left[\bar\psi\left(\sigma_{\rm av}+i\vec{\tau}\cdot\vec{\pi}_{\rm av}
\gamma_5\right)\psi\right] \right. \nonumber \\  
\left. - {\partial \sigma_{\rm in}\over \partial r} {\partial\sigma_{\rm out} 
\over \partial t} + {\partial\sigma_{\rm out}\over \partial r} 
{\partial\sigma_{\rm in} \over \partial t} \right. - \nonumber \\
\left. {\partial\pi_{\rm in}\over 
\partial r} {\partial\pi_{\rm out} \over \partial t} +
{\partial\pi_{\rm out} \over \partial r} {\partial\pi_{\rm in} \over 
\partial t} \right\} \; .
\la{rdoteq}
\eea
We use the above expression to compute the motion of the bag's surface, 
which conserves the total energy and momentum.

It is important to notice that for spherically symmetric solutions the 
total linear momentum is conserved regardless of the value of $B$, 
because the associated current is radial and the vector sum
always gives a zero total momentum. This guarantees the 
conservation of the total momentum also for a
non-static bag surface, because in such a case the RHS of 
Eq.~\ref{bagconstant} is not constant and
hence the equation is not satisfied.

The first question we want to address is whether the static hedgehog 
solution is stable with respect to a 
small perturbation or it is just a special field configuration  
permitted only with a static boundary.
If the static hedgehog models a hadron state one would like it to be 
little 
affected by a small non-resonant perturbation. We therefore considered 
an incoming wavepacket incident on the bag, and
we computed the evolution of the system. We used both a $\pi$-field 
and a $\sigma$-field as the incoming packet, with the following form
\be
\begin{array}{lll}
\left . \begin{array}{l}
\sigma_{\rm out}(t_0,r) \\
\pi_{\rm out}(t_0,r) 
\end{array} \right \} & = &
A \left[e^{-\beta(r-R_0)}-1\right]^3 \cdot \\
& & \sin\left(\nu (t_0+r)\right) 
e^{-\alpha(r-R_0)} \hspace{0.8 cm} r > R_0 \, ,
\end{array}
\la{wvpacket}
\ee
which, at $r=R_0$, is zero up to the third derivative, in order to avoid 
discontinuities at the instant of the collision.
The motion of the bag surface depends on the bag constant $B$. 
We use $B=1$ fm$^{-4}$ to demonstrate the qualitative features of the system.

\begin{figure}
\psfig{file=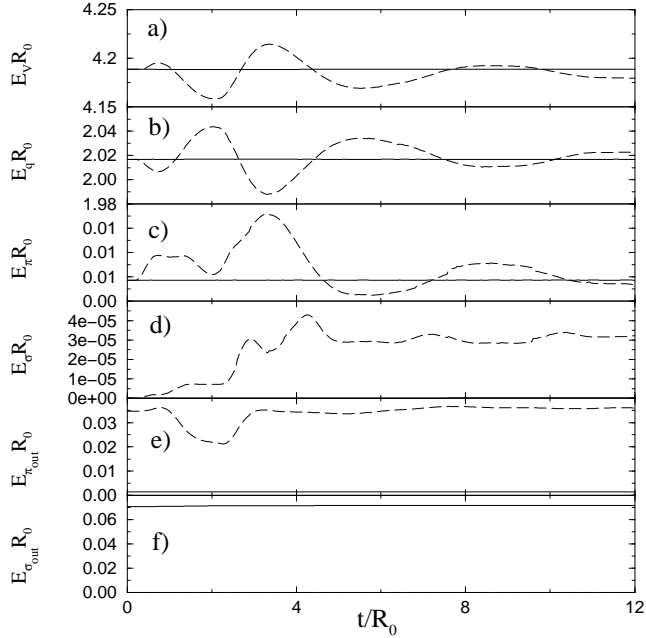,angle=-90,width=8.5 cm}

\caption{Time evolution of the energy in the case of a 
collision with a wavepacket with $\nu=1/R_0$, 
$\alpha=0.2/R_0$, $\pi$-wave with $A=0.05/R_0$ (dashed line) or
$\sigma$-wave with $A=0.1/R_0$ (solid line) (see Eq.~\ref{wvpacket}):
a) volume energy, b) fermion energy,
c) $\pi$-field energy inside the bag, d) $\sigma$-field energy inside the 
bag,
e) $\pi$-field energy outside the bag, and
f) $\sigma$-field energy outside the bag.}
\label{fig6}
\end{figure}

In Fig.~\ref{fig6} we show the energy of the
fields inside and outside the bag versus time for small $A$ and $\nu$. In all
the computations we used $\beta=1/R_0$.
Both for a $\pi$-wave and
a $\sigma$-wave the bag is hardly changed, and after the interaction 
the velocity of the surface
goes back to zero gradually as expected. Part of the pion field is 
reflected back at the surface of the bag,
while the other part after penetrating the bag goes back out towards 
infinity.
It is interesting to observe that a $\sigma$-wave has almost no effect at 
all on the bag and nearly does not enter it. 
We verified that the static hedgehog solution remains little affected also 
for larger $A$ and higher frequencies $\nu$, thus
validating its use as a stationary state of a hadron.

\subsection{Resonances}

We next consider whether the resonances found in 
the case of a driven bag motion still
occur for the self-consistent motion caused by an incoming wave of 
appropriate frequency. This is a non-trivial
question because the nonlinear relation between the fields and the motion 
of the boundary, as expressed in 
Eq.~\ref{rdoteq}, might in principle destroy any phase coherence on which 
a resonance is built up.
We hence performed our computation with incoming $\pi$-waves given by  
Eq.~\ref{wvpacket} with $\alpha = 0$
and $\nu = n\pi/R_0$, $\nu = (E_n - E_k)$, and $\nu = 
(2n+1)\pi/(2R_0)$. Again due to numerical limitations
we had to use small values of $A$.

\begin{figure}
\psfig{file=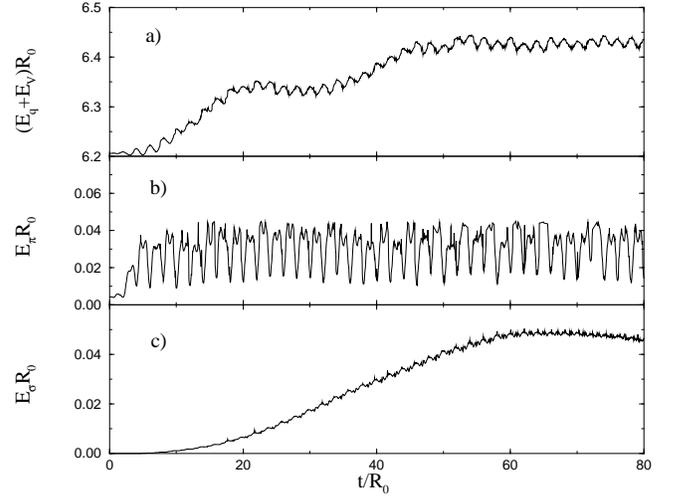,angle=-90,width=8.5 cm}

\caption{Time evolution of the energy for a resonance 
with an incoming $\nu=\pi/R_0$, $A=0.005/R_0$, $\alpha=0$ $\pi$-wave 
(see Eq.~\ref{wvpacket}), showing a) fermion energy plus volume energy, b) 
$\pi$-field energy
inside the bag, and c) $\sigma$-field energy inside the bag.}
\label{fig8}
\end{figure}

\begin{figure}
\psfig{file=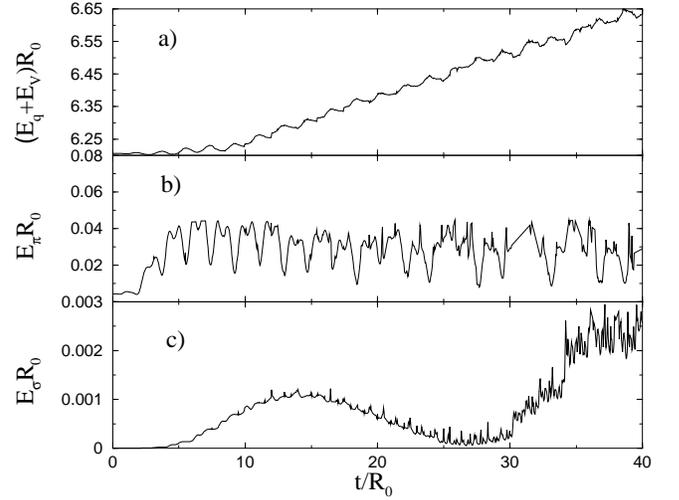,angle=0,width=8.5 cm}

\caption{Same as Fig.~6, but for $\nu=E_2-E_1$.}  
\label{fig9}
\end{figure}

In Fig.~\ref{fig8} we plot the energy of the fields inside the bag 
vs.~time for $\nu=\pi/R_0$ and in Fig.~\ref{fig9}
for  $\nu=E_2 - E_1$. We can see that the resonant behaviour is still 
present. From the point of view of the 
energy gained by the bag the two resonances merge and appear as a broad 
resonance peaked at $\nu= E_2-E_1$. 
With a closer look however one can observe two different physical 
phenomena. For $\nu$ close to $\pi/R_0$ the
$\sigma$-field is excited and the bag expands while the fermion field 
gets only a small contribution
from the second static hedgehog state. For $\nu$ close to $E_2 - E_1$ the 
$\sigma$-field is slightly out of 
resonance while the fermion field tends to be excited towards the second 
static hedgehog state and the volume of the bag decreases.

\begin{figure}
\psfig{file=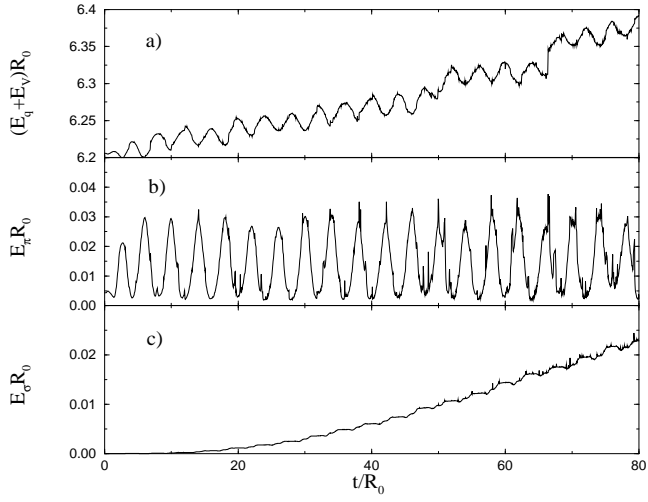,angle=-90,width=8.5 cm}

\caption{Same as Fig.~6, but for $\nu=\pi/(2R_0)$ and $A=0.02/R_0$.} 
\label{fig10}
\end{figure}

In the case of a wave with $\nu = \pi/(2R_0)$ we also have a clear 
resonance that involves the $\sigma$-field (Fig.~\ref{fig10}). At this
frequency the fermion field is completely out of resonance. Overall the 
bag's energy increases
not only as the pion energy but also in the form of volume energy due 
to the expansion of the bag. 
The excitation mechanism for the $\sigma$-field is the same as explained 
in the previous section by means
of Eqs.~\ref{pioq}~and~\ref{sigmaoq}.

It is very interesting to notice that the expansion of the bag is related 
to the excitation of the $\sigma$-field and not directly to the fermions.
The increase of energy due to a larger bag radius
is the classical counterpart of the breathing modes proposed by other 
authors \cite{rebbi,kuti,brown,meissner,nogami} to explain certain radial
excitations  of the baryons such as the Roper resonance. These authors propose
that  such resonances are excitations
of the collective degrees of freedom of the bag, represented in the 
models they considered by the surface coordinates,
 while the quarks essentially remain in the ground state. In the 
Chiral Bag model, besides the bag's radius, 
the pions too describe collective degrees of freedom, and hence our 
results strongly support the above scenario.

For larger odd multiples of $\pi/(2R_0)$ the resonant behaviour is much 
attenuated. We believe that this is due to
the fact that at higher frequencies and with a self-consistent bag 
motion an approximation as the one shown in 
Eqs.~\ref{gst}~and~\ref{fst} is no longer acceptable. The perturbation 
still appears to be resonant for the $\sigma$ 
field, but its energy increases very slowly.

\begin{figure}
\psfig{file=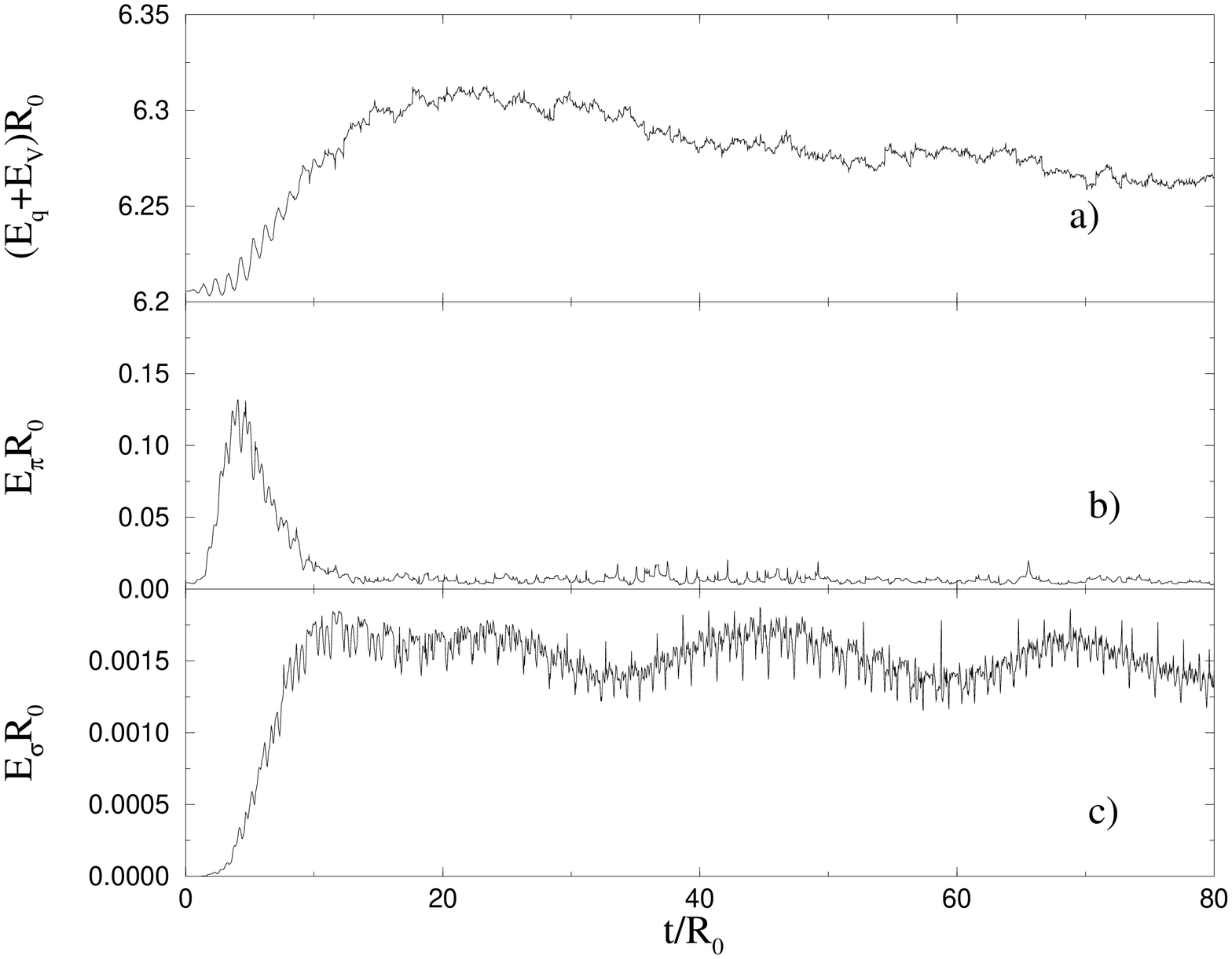,angle=-90,width=8.5 cm}

\caption{Same as Fig.~6, but for $\nu=6.55/R_0 \simeq E_3-E_1$, $A=0.005/R_0$ 
and $\alpha=0.2/R_0$.}
\label{fig11}
\end{figure}

Another remarkable property of the chiral bag is that it shows a 
realistic behaviour in a scattering process.
We have already mentioned the dynamics for the scattering with a 
non-resonant pion wavepacket, but it is interesting to
examine how the bag releases its energy after being excited by a resonant 
wavepacket. In Figs.~\ref{fig11} it
can be seen how, after the incoming wavepacket has been scattered away, 
the bag remains in its excited
state for some time before starting to  slowly release the energy gained, 
hence looking like a stable particle. 
Also in this case we can see that 
the $\pi$-wave inside the bag is not excited and its temporary increase 
in energy is simply due to the part of the incoming wave
that enters the bag before being reflected away.

Since the chiral bag is generally used to describe baryons, we have also 
performed the previous calculations with
three quarks inside the bag in order to verify that our findings hold in 
this case too. Not having quantized the theory, 
we have to consider already in the Lagrangian three distinct fermion fields. 
Moreover, the coupling with the pions causes the solution 
to differ from the one-quark bag. In fact, while each quark has to 
satisfy the same 
equations Eqs.~\ref{eqmn1}~and~\ref{eqmn2}, the equations for the pions 
change because the RHS of 
Eqs.~\ref{eqmn3}~and~\ref{eqmn4} must be multiplied by a factor $3$. Such 
a difference produces different eigenvalues
for the static cavity solution and yields the following equation for the 
motion of the bag's surface
\be
\begin{array}{lll}
\dot{R} & = & {-1 \over B-D(t)}\left\{ \hspace{-0.2 cm} \begin{array}{l}
\\
\\
\end{array}
{3\over 2f_\pi}{\partial \over \partial 
t}\left[\bar\psi\left(\sigma_{\rm av}+i\vec{\tau}\cdot\vec{\pi}_{\rm av}
\gamma_5\right)\psi\right] \right. \\
& & \\
& & \left. - {\partial \sigma_{\rm in}\over \partial r} 
{\partial\sigma_{\rm out} \over \partial t} + 
{\partial\sigma_{\rm out}\over \partial r} 
{\partial\sigma_{\rm in} \over \partial t} \right. \\
& & \\
& & \left. - {\partial\pi_{\rm in}\over 
\partial r} {\partial\pi_{\rm out} \over \partial t} +
{\partial\pi_{\rm out} \over \partial r} {\partial\pi_{\rm in} \over 
\partial t} \hspace{-0.2 cm} \begin{array}{l}
\\
\\
\end{array}
\right\} \; .
\end{array}
\la{rdoteq3q}
\ee
We have 
verified that the features found with only one quark
remain with three quarks.

For smaller values of $B$ or bigger amplitudes $A$ we have been
able to obtain fairly accurate solutions for short times ($t < 10$ fm$/c$) 
and, for resonant perturbations, we observe a much stronger and faster
excitation process, which could probably model realistic energy levels.

\section{Similarities with the soliton} 
 
We point out here the very interesting similarities 
between the hedgehog solution and the soliton solution of a non-linear 
field theory. 
 
We can define a classical soliton as {\it any spatially confined and  
nondispersive solution of a classical field theory} \cite{tdlee}. 
In order  to have soliton solutions it is necessary to have some nonlinear  
couplings among the fields. The MIT bag model with only 
fermions inside the bag does not have nonlinear  
couplings. In fact, as we proved in a previous paper \cite{colchu3}, it 
admits bag-like solutions, but these are unstable with respect to 
perturbations of the bag surface. 
In the chiral bag model the quark-pion coupling,  
although it is linear, introduces a nonlinear self-coupling  
for the fermion field through the boundary conditions, as apparent 
from Eqs.~\ref{bndsigma},~\ref{bndpi}. We have seen that the hedgehog 
solution is indeed stable with respect to perturbations of the bag surface. 
 
For a boson field the various nonlinear couplings can be  
characterized by a dimensionless coupling constant $g$. 
If $g=0$, the theory is linear and there is no soliton solution. 
However if $g$, however small, is different from zero, the theory 
admits soliton solutions. In the limit $g \rightarrow 0$ the soliton 
solution grows to infinity. This is remarkably similar to what  
happens in the dynamical chiral bag model. The dimensionless coupling 
constant is in this case $\gamma/f_\pi$, with $\gamma$ an arbitrary constant 
with dimension of $L^{-1}$. If we set $\gamma/f_\pi = 0$ already in the  
Lagrangian, we have the MIT bag model, and no stable, spatially 
confined solution exists. 
If we take the limit $\gamma/f_\pi \rightarrow 0$ 
($f_\pi \rightarrow \infty$), 
we still have stable hedgehog solutions, but the field $\sigma$ goes 
to infinity, so that $\sigma/f_\pi \rightarrow 1$. 
 
Such similarities seem more than a coincidence, especially if we consider 
the bag models as simplifications of more general models. In fact 
it has been shown \cite{gomm} that a chiral model, similar to the 
Skyrme Lagrangian, can automatically produce bag-like solutions. 
From this point of view one may put some features of the bag-like 
solutions already in the Lagrangian, thus obtaining a bag model. 
The fact that the MIT bag model does not admit a stable bag-like  
solution may be viewed as due to an oversimplification, having  
completely neglected the quark-pion interaction, while in the 
chiral bag model such interaction is maintained at least  
at the surface of the bag. 
 
\section{Conclusion}

We have shown that the chiral bag described by the 
Lagrangian Eq.~\ref{lagct} is stable
in the sense that a bag-like solution exists even if the static bag is 
perturbed either by arbitrary radial motions of
its boundary or by its interaction with a meson wave. 
This is in contrast with the purely fermionic MIT bag which
has been proved to be unstable \cite{colchu3}.

Computing the solution for a bag perturbed by a non-resonant meson wave, 
we found that it remains close to
the static hedgehog solution and that, after the incoming wave is 
scattered away, it returns to the static
hedgehog. Such a result validates the use of the static hedgehog as a 
stationary state of a hadron.

We also examined the existence of resonant perturbations, and we 
discovered three kinds of resonances which occur
when the bag interacts with an incoming $\pi$-wave. When the frequency 
of the incoming wave is close to an energy gap,
$\nu \simeq E_n-E_k$, the fermion field is in resonance.
The $\sigma$-field is also excited since $\nu$ is close to
an integral multiple of $\pi/R_0$. The fermion field tends to go to an upper
static hedgehog level, but these resonances
are not simply transitions from a lower hedgehog state to an upper one, 
because in that case there would be only
a static meson field in the final state, while here we have remarkable 
energy contribution from a non-static $\sigma$-field.
At $\nu=n\pi/R_0$ the fermion field is little excited while the 
$\sigma$-field is in resonance and the bag expands.
The third type of resonance occurs when $\nu$ equals odd multiples of 
$\pi/(2 R_0)$,  
which is a consequence of the linear boundary condition Eq.~\ref{eqmn2}. 
In this case the fermion field is 
not excited and the increase in energy comes from the $\sigma$-field and 
the increase in the volume energy associated with the expansion of the bag. 
The occurence of a resonance at $\nu=\pi/(2 R_0)$, which is much smaller than
the energy gap between the first and second static hedgehog states, gives 
support to the description 
of the Roper resonance as a radial excitation of the collective degrees 
of freedom. In particular the expansion of
the bag without fermion excitation is the classical analog of the 
breathing modes proposed by other authors 
\cite{rebbi,kuti,brown,meissner,nogami}.
Our results, however, show that the expansion of the bag is strictly 
related to the excitation of the $\sigma$-field and in this sense 
warn against the use of the adiabatic approximation. 

Compared to previous studies of dynamical bag models, 
our approach has two main advantages: we solve the equations of motion 
without approximations and we show the dynamics of the resonances.

In the present work we have considered hedgehog configurations in 
which the quarks are neither in
a flavor eigenstate nor in a $J$-eigenstate; they do not represent any 
known hadrons. However the occurence
of the resonances we found does not depend on the flavor of the quarks, 
and since they are caused by 
spherical waves the angular momenta are not changed. Moreover they 
are not related to specific features
of the hedgehog solution. In fact the $\nu=E_n-E_k$ resonance is a 
general feature of discrete-level 
systems. The resonances at $\nu=n\pi/R_0$ are geometrical resonances 
related to the fact that the mesons in this model
are massless, and the ones at $\nu=(2n+1)\pi/(2 R_0)$ are related to the 
linear boundary condition Eq.~\ref{eqmn2}.
Therefore, we conjecture that such resonances should occur also for more 
realistic solutions, {\it i.e.}~with definite flavor and $J$.

We thank the support of a Hong Kong Research Grants Council grant CUHK
4189/97P and a Chinese University Direct Grant (Grant No.~2060193).

\section{Appendix}

For a spherical bag $n_\mu$ can be written as $(\rp , -\hat{r})$ and, as 
shown for example in \cite{colchu3}, 
Eq.~\ref{eqmn2} in radial coordinates becomes

\be
i\dot{R} g(t,R) - f(t,R) = {1\over f_\pi}\left[\sigma(t,R) g(t,R) - 
\pi(t,R) 
f(t,R) \right] ,
\la{bnddirspin1}
\ee
\be
-i\dot{R} f(t,R) - g(t,R) = {1\over f_\pi}\left[\sigma(t,R) f(t,R) + 
\pi(t,R) 
g(t,R) \right] .
\la{bnddirspin2}
\ee
Equating separately the real and imaginary parts of the two equations we 
obtain the following four equations

\be
\begin{array}{l}
-\dot{R} g_{\rm im}(t,R) - f_{\rm re}(t,R) = \\
\\
{1 \over f_\pi}\left[
\sigma(t,R) g_{\rm re}(t,R) - \pi(t,R) f_{\rm re}(t,R)\right]\ \ ,
\end{array}
\la{bnddirspinreim1}
\ee
\be
\begin{array}{l}
\dot{R} f_{\rm im}(t,R) - g_{\rm re}(t,R) = \\
\\
{1 \over f_\pi}\left[
\sigma(t,R) f_{\rm re}(t,R) + \pi(t,R) g_{\rm re}(t,R)\right]\ \ ,
\end{array}
\la{bnddirspinreim2}
\ee
\be
\begin{array}{l}
\dot{R} g_{\rm re}(t,R) - f_{\rm im}(t,R) =  \\
\\
{1 \over f_\pi}\left[
\sigma(t,R) g_{\rm im}(t,R) - \pi(t,R) f_{\rm im}(t,R) \right]\ \ ,
\end{array}
\la{bnddirspinreim3}
\ee
\be
\begin{array}{l}
-\dot{R} f_{\rm re}(t,R) - g_{\rm im}(t,R) = \\
\\
{1 \over f_\pi}\left[
\sigma(t,R) f_{\rm im}(t,R) + \pi(t,R) g_{\rm im}(t,R) \right] \ \ .
\end{array}
\la{bnddirspinreim4}
\ee

At this point we have four equations and four unknowns, {\it i.e.}~ 
$Q_{\rm re}(t+R)$, $Q_{\rm im}(t+R)$, 
$\sigma(t,R)$, and $\pi(t,R)$. This fact seems to make our requirement, 
that $\sigma(t,r)$ and $\pi(t,r)$ be continuous at $r=R$, redundant hence 
making the whole problem inconsistent. In fact if we could derive from 
Eqs.~\ref{bnddirspinreim1}~,~\ref{bnddirspinreim2}~,
~\ref{bnddirspinreim3}~and~\ref{bnddirspinreim4}
all the four functions mentioned above,  
then Eqs.~\ref{eqmn3rd}~and~\ref{eqmn4rd} would require 
$\sigma_{\rm out}$ and $\pi_{\rm out}$ to be
singular at $r=R$, and even this would not guarantee the 
existence of a solution in general.
However, it turns out that 
Eqs.~\ref{bnddirspinreim1}~,~\ref{bnddirspinreim2}~,
~\ref{bnddirspinreim3}~and~\ref{bnddirspinreim4}
are not independent and we need to 
impose some condition on the functions in order to have a 
unique solution. We have verified this in two ways, as explained below.

Solving 
Eqs.~\ref{bnddirspinreim1},~\ref{bnddirspinreim2},~\ref{bnddirspinreim3}~and~\ref{bnddirspinreim4}
for $\sigma(t,R)$ and $\pi(t,R)$ we 
obtain two different expressions
for each function

\be
\begin{array}{l}
\sigma(t,R)=f_{\pi} \left\{ \hspace{-0.2 cm} \begin{array}{l}
\\
\\
\end{array}
-2 f_{\rm im}(t,R) g_{\rm im}(t,R) + [g_{\rm re}(t,R) 
g_{\rm im}(t,R) \right. - \\
\\
\left. f_{\rm re}(t,R) f_{\rm im}(t,R)]\dot{R} \hspace{-0.2 cm} \begin{array}{l}
\\
\\
\end{array}
\right\} /
\left( g_{\rm im}^2(t,R) + f_{\rm im}^2(t,R) \right) \: ,
\end{array}
\la{sigma1}
\ee
\be
\begin{array}{l}
\sigma(t,R)=f_{\pi}\left\{ \hspace{-0.2 cm} \begin{array}{l}
\\
\\
\end{array}
-2 f_{\rm re}(t,R) g_{\rm re}(t,R) - [g_{\rm re}(t,R) 
g_{\rm im}(t,R) \right. - \\
\\
\left. f_{\rm re}(t,R) f_{\rm im}(t,R)]\dot{R} \hspace{-0.2 cm} \begin{array}{l}
\\
\\
\end{array}
\right\} /
\left( g_{\rm re}^2(t,R) + f_{\rm re}^2(t,R)\right) \: ,
\end{array}
\la{sigma2}
\ee
\be
\begin{array}{l}
\pi(t,R)=f_{\pi}\left\{\hspace{-0.2 cm} \begin{array}{l}
\\
\\
\end{array}
 f_{\rm im}^2(t,R) - g_{\rm im}^2(t,R) - [f_{\rm re}(t,R) 
g_{\rm im}(t,R) + \right. \\
\\
\left. g_{\rm re}(t,R) f_{\rm im}(t,R)]\dot{R} \hspace{-0.2 cm} \begin{array}{l}
\\
\\
\end{array}
\right\}/
\left( g_{\rm im}^2(t,R) + f_{\rm im}^2(t,R)\right) \: ,
\end{array}
\la{pi1}
\ee
\be
\begin{array}{l}
\pi(t,R)=f_{\pi}\left\{\hspace{-0.2 cm} \begin{array}{l}
\\
\\
\end{array}
 f_{\rm re}^2(t,R) - g_{\rm re}^2(t,R) + [f_{\rm re}(t,R) 
g_{\rm im}(t,R) + \right. \\
\\
\left. g_{\rm re}(t,R) f_{\rm im}(t,R)]\dot{R} \hspace{-0.2 cm} \begin{array}{l}
\\
\\
\end{array}
\right\}/
\left( g_{\rm re}^2(t,R) + f_{\rm re}^2(t,R)\right) \: .
\end{array}
\la{pi2}
\ee
Equating Eq.~\ref{sigma1} with Eq.~\ref{sigma2} and 
Eq.~\ref{pi1} with Eq.~\ref{pi2}
we obtain two non-linear equations for 
the real and imaginary parts of $f$ and $g$,

\be
\begin{array}{l}
\left\{ \hspace{-0.2 cm} \begin{array}{l}
\\
\\
\end{array}
2 f_{\rm im}(t,R) g_{\rm im}(t,R) - \left[g_{\rm re}(t,R) 
g_{\rm im}(t,R) \right. \right. - \\
\\
\left. \left. f_{\rm re}(t,R) f_{\rm im}(t,R)\right] \dot{R} \hspace{-0.2 cm} \begin{array}{l}
\\
\\
\end{array}
\right\} 
\cdot \left[g_{\rm re}^2(t,R) + f_{\rm re}^2(t,R)\right] = 
\\
\\
\left\{ \hspace{-0.2 cm} \begin{array}{l}
\\
\\
\end{array}
2 f_{\rm re}(t,R) g_{\rm re}(t,R) + \left[g_{\rm re}(t,R) 
g_{\rm im}(t,R) \right. \right. - \\
\\
\left. \left. f_{\rm re}(t,R) f_{\rm im}(t,R)\right] \dot{R} \hspace{-0.2 cm} \begin{array}{l}
\\
\\
\end{array}
\right\} 
\cdot \left[g_{\rm im}^2(t,R) + f_{\rm im}^2(t,R)\right]  \: ,
\end{array}
\la{qimqreeq1}
\ee

\be
\begin{array}{l}
\left\{ \hspace{-0.2 cm} \begin{array}{l}
\\
\\
\end{array}
f_{\rm im}^2(t,R) - g_{\rm im}^2(t,R) - \left[f_{\rm re}(t,R) 
g_{\rm im}(t,R) \right. \right. + \\
\\
\left. \left. g_{\rm re}(t,R) f_{\rm im}(t,R)\right]\dot{R} \hspace{-0.2 cm} \begin{array}{l}
\\
\\
\end{array}
\right\} 
\cdot \left[g_{\rm re}^2(t,R) + f_{\rm re}^2(t,R)\right] =  
\\
\\
\left\{ \hspace{-0.2 cm} \begin{array}{l}
\\
\\
\end{array}
f_{\rm re}^2(t,R) - g_{\rm re}^2(t,R) + \left[f_{\rm re}(t,R) 
g_{\rm im}(t,R) \right. \right. + \\
\\
\left. \left. g_{\rm re}(t,R) f_{\rm im}(t,R)\right]\dot{R} \hspace{-0.2 cm} \begin{array}{l}
\\
\\
\end{array}
\right\} 
\cdot \left[g_{\rm im}^2(t,R) + f_{\rm im}^2(t,R)\right] \: .
\end{array}
\la{qimqreeq2}
\ee

The problem is evidently extremely difficult to handle analytically, and 
so we used a numerical approach.
Substituting $Q^\prime$ with a finite incremental ratio, 
Eqs.~\ref{qimqreeq1}~and~\ref{qimqreeq2} become two 
non-linear algebraic equations for $Q_{\rm re}(t+R)$ and $Q_{\rm im}(t+R)$. 
Solving numerically the algebraic equations,
we have found that a whole 
region exists, in the $Q_{\rm re}$ --- $Q_{\rm im}$ plane around 
$Q(t+R-dz)$, in which the algebraic equations are satisfied, 
thus indicating that the solution is not unique.
Since this is not a rigorous proof, we need to cross-check our finding 
by imposing the continuity of $\sigma(t,r)$ and
$\pi(t,r)$ on the surface of the bag and by verifying whether the 
solutions, found without using 
Eqs.~\ref{qimqreeq1}~and~\ref{qimqreeq2}, satisfy all the four 
Eqs.~\ref{bnddirspinreim1},~\ref{bnddirspinreim2},~\ref{bnddirspinreim3}~and~\ref{bnddirspinreim4}.

In order to do this we have solved numerically  
Eqs.~\ref{bndsigma}~and~\ref{bndpi} for $Q_{\rm re}(t+R)$ and 
$Q_{\rm im}(t+R)$, as discussed in the main part of the paper,
with $\sigma(t,R)$ and $\pi(t,R)$ 
given by anyone of the above expressions or a combination of them. 
We thereby verified that 
Eqs.~\ref{bnddirspinreim1},~\ref{bnddirspinreim2},~\ref{bnddirspinreim3},~\ref{bnddirspinreim4} 
are automatically satisfied.

In the case of $\rp = 0$ we can analytically prove that 
Eqs.~\ref{bnddirspinreim1},~\ref{bnddirspinreim2},
~\ref{bnddirspinreim3}~and~\ref{bnddirspinreim4}
are not independent. In fact, looking for solutions of the
form $g_{re}=P(t) g(r)$,  $f_{re}=P(t) f(r)$, $g_{im}=S(t) g(r)$,
$f_{im}=S(t) f(r)$, we easily verify that Eqs.~\ref{qimqreeq1},~\ref{qimqreeq2}
are automatically satisfied leaving $P(t)$ and $S(t)$ undetermined.
It is reasonable to think that as $\rp$ slowly departs from zero
the four equations still remain not independent, though we are not
able to provide a rigorous proof for it.

%

%

%

\end{multicols}


\begin{thebibliography}{33}

\bibitem{MIT} A.~Chodos, R.~L.~Jaffe, K.~Johnson, C.~B.~Thorn and
V.~F.~Weisskopf, Phys.~Rev.~D {\bf 9} (1974) 3471.

\bibitem{chodos}A.~Chodos and C.~B.~Thorn, Phys.~Rev.~D {\bf 12}, 2733 (1975).

\bibitem{chbag} See for example, \\
F.~Myhrer, in {\it Quarks and Nuclei},
Int.~Rev.~Nucl.~Phys.~{\bf 1} (1984) 326. (Ed.~W.~Weise,
World Scientific, Singapore, 1984);
G.~A.~Miller, in {\it Quarks and Nuclei},
Int.~Rev.~Nucl.~Phys.~{\bf 1} (1984) 326. (Ed.~W.~Weise,
World Scientific, Singapore, 1984); S.~Th\'eberge, A.~W.~Thomas and
G.~A.~Miller, Phys.~Rev.~D {\bf 22} (1980) 2838, D {\bf 23} (1981);
2106(E), D {\bf 24} (1981) 216; A.~W.~Thomas, Advances in 
Nucl.~Phys.~{\bf 13}
(1983) 1, Eds.~J.~Negele and E.~Vogt, Plenum Press, N.Y.

\bibitem{cloudy}S.~Th\'eberge, A.~W.~Thomas and G.~A.~Miller,
Phys.~Rev.~D {\bf 22}, 2838 (1980);
A.~W.~Thomas, Advances in Nucl.Phys., {\bf 13}, 1 (1983),
Eds.~J.~Negele and E.~Vogt, Plenum Press, N.Y..

\bibitem{sstar} E.~Witten, Phys.~Rev.~D {\bf30} (1984) 272;
A.~R.~Bodmer, Phys.~Rev.~D {\bf 4} (1971) 1601;
P.~Haensel, J.~L.~Zdunik and R.~Schaeffer,
Astron.~Astrophys.~{\bf 160} (1986) 121; C.~Alcock, E.~Farhi and
A.~V.~Olinto, Ap.~J.~{\bf 310} (1986) 261.

\bibitem{rhic} See for example, L.~McLerran, Rev.~Mod.~Phys. {\bf 58}, 
1021 (1986). 


\bibitem{rebbi} C.~Rebbi, Phys.~Rev.~D {\bf 12}, 2407 (1975);
C.~Rebbi, Phys.~Rev.~D {\bf 14}, 2362 (1976);
T.~DeGrand and C.~Rebbi, Phys.~Rev.~D {\bf 17}, 2358 (1978).

\bibitem{kuti} P.~Hasenfratz and J.~Kuti,
Phys.~Rep.~{\bf 40}, 75 (1978).

\bibitem{brown} G.~E.~Brown, J.~W.~Durso and M.~B.~Johnson,
Nucl.~Phys.~A {\bf 397}, 447 (1983).

\bibitem{meissner} U.~-G.~Meissner and J.~W.~Durso,
Nucl.~Phys.~A {\bf 430}, 670 (1984).

\bibitem{nogami} Y.~Nogami and L.~Tomio, Can.~J.~Phys.~{\bf 62}, 260 (1984); 
Y.~Nogami and L.~Tomio, Phys.~Rev.~D {\bf 31}, 2818 (1985).

\bibitem{colchu3}K.~Colanero and M.~-C.~Chu, submitted to J.~Phys.~A.



\bibitem{bhaduri} R.~K.~Bhaduri, {\it Models of the nucleon}
(Addison-Wesley, Redwood City, 1988).

\bibitem{cole}C.~K.~Cole and W.~C.~Schieve, Phys.~Rev.~A {\bf 52}, 
4405 (1995), and references therein.

\bibitem{colchu2}K.~Colanero and M.~-C.~Chu, Phys.~Rev.~E {\bf 62}, 8663 
(2000) and references therein.


\bibitem{colchu1}K.~Colanero and M.~-C.~Chu, Phys.~Rev.~A {\bf 60 }, 1845 
(1999).

\bibitem{wai}K.~W.~Chan, U.~M.~Ho, P.~T.~Leung, and M.-C.~Chu, 
The Chinese University of Hong Kong Preprint, 2000 (unpublished); 
K.~W.~Chan, Master Thesis, The Chinese University of Hong Kong
(unpublished), 1999.


\bibitem{tdlee}T.~D.~Lee, {\it Particle Physics and Introduction to 
Field Theory} (Harwood Academic Publishers 1988). 

\bibitem{gomm}H.~Gomm, P.~Jain, R.~Johnson and J.~Schechter, 
Phys.~Rev.~D {\bf 33}, 3476 (1986). 


\end{thebibliography}
\end{document}